\documentclass[fleqn,10pt]{wlscirep}
\usepackage[utf8]{inputenc}
\usepackage[T1]{fontenc}
\usepackage{parskip}
\usepackage{textcomp, gensymb}
\usepackage[
backend=biber,
style=numeric-comp,
sorting=none
]{biblatex}
\usepackage{csquotes}
\usepackage{xurl}

\title{Detection of the infrared aurora at Uranus with Keck-NIRSPEC}

\author[1,*]{Emma M. Thomas}
\author[1]{Henrik Melin}
\author[2]{Tom S. Stallard}
\author[1]{Mohammad N. Chowdhury}
\author[1]{Ruoyan Wang}
\author[2]{Katie Knowles}
\author[3]{Steve Miller}
\affil[1]{School of Physics and Astronomy, University of Leicester, Leicester, UK}
\affil[2]{Department of Mathematics, Physics and Electrical Engineering, Northumbria University, Newcastle upon Tyne, UK}
\affil[3]{Department of Physics and Astronomy, University College London, London, UK}

\affil[*]{emt18@leicester.ac.uk}

\begin{abstract}
Near-infrared (NIR)-wavelength observations of Uranus have been unable to locate any infrared aurorae, despite many attempts to do so since the 1990s. While at Jupiter and Saturn, NIR investigations have redefined our understanding of magnetosphere–ionosphere–thermosphere coupling, the lack of NIR auroral detection at Uranus means that we have lacked a window through which to study these processes at Uranus. Here we present NIR Uranian observations with the Keck II telescope taken on the 5 September 2006 and detect enhanced $\text{H}_{\text{3}}^{\text{+}}$ emissions. Analysing temperatures and column densities, we identify an 88\% increase in localized $\text{H}_{\text{3}}^{\text{+}}$ column density, with no significant temperature increases, consistent with auroral activity generating increased ionization. By comparing these structures against the $\text{Q}_{\text{3}}^{\text{mp}}$ magnetic-field model and the Voyager 2 ultraviolet observations, we suggest that these regions make up sections of the northern aurora.
\end{abstract}
\begin{document}

\maketitle
\thispagestyle{empty}

\section*{Main}

One of the most unusual aspects of Uranus is its magnetic field, off centred by a third of the planet’s radius and tilted 59\degree from the planet’s rotational axis \cite{Ness:1986}. With the planetary rotational axis also tilted by 98\degree, we are presented with a truly distinctive magnetosphere. The only flyby of the planet was made by Voyager II in 1986, where the data presented more questions than answers. Aurorae, presumably created by interactions between Uranus’s magnetosphere and ionosphere, were first simultaneously detected by several instruments onboard Voyager II. Both radio (UKR) and ultraviolet (UV) emissions have shown auroral morphologies that are very different from those seen at Jupiter and Saturn and provide the foundation for auroral emission comparisons \cite{Curtis:1987},\cite{Kaiser:1987},\cite{Herbert:2009}. Investigations into Uranus’s aurora have continued with the Hubble Space Telescope (HST) \cite{Lamy:2012},\cite{Lamy:2017},\cite{Lamy:2018},\cite{Lamy:2020}. In these studies, 15 UV auroral emissions were identified: 9 southern aurora and 6 northern aurora emissions, largely consistent with pulsed cusp aurorae and observed to emit conjugately. In these previous investigations, the solar-wind conditions were found to greatly affect the Uranian aurorae, where Voyager II revealed a correlation between active solar-wind periods and UKR activity \cite{Desch:1989}, with HST data showing positive detections (~25\%) of UV auroral emissions that coincided with modelled peaks in solar-wind activity \cite{Lamy:2012},\cite{Lamy:2017},\cite{Lamy:2018},\cite{Lamy:2020}. It is noted that one UKR component (known as n-smooth) observed by \cite{Kaiser:1989}, originates from close to the magnetic equator at 2–3 $\text{R}_{\text{U}}$ and was hence not considered auroral.

NIR emissions from $\text{H}_{\text{3}}^{\text{+}}$ (a molecular ion) have been fundamental in developing our understanding of the aurorae at Jupiter and Saturn \cite{Drossart:1989, Geballe:1993, Stallard:2008, Stallard:2012, Stallard:2019, Johnson:2017, Johnson:2018, Chowdhury:2019, Miller:2000, Miller:2006, Miller:2020}. $\text{H}_{\text{3}}^{\text{+}}$ was first discovered at Uranus in 1992 (\cite{Trafton:1993}), and has been frequently analysed to characterize the ionosphere and understand seasonal and temporal changes therein. For 30 years, a continuous effort was made to document an infrared aurorae at Uranus \cite{Stallard:2012},\cite{Lam:1997},\cite{Trafton:1999},\cite{Melin:2011},\cite{Melin:2019}. In Ref \cite{Lam:1997}, a 20\% variation in $\text{H}_{\text{3}}^{\text{+}}$ emissions was tentatively attributed to auroral processes but due to the low signal-to-noise ratio, it could not be confirmed whether these increases were auroral. In \cite{Melin:2019}, a localized $\text{H}_{\text{3}}^{\text{+}}$ emission peak was observed on the dawnside limb of Uranus. Its location aligned with the southern aurorae latitudes; however, owing to time constraints, the feature could not be tracked and remains unconfirmed. From here on, when discussing the geometry of Uranus, we refer to the Uranian longitude coordinate system (ULS) presented by \cite{Ness:1986}. In addition, the exact longitude of Uranus during observations is unknown owing to the ±0.01 h rotational period uncertainty; hence, the longitude at Uranus is completely lost in ~3.4 Earth years.

In this Article, we present high-resolution IR emissions at Uranus obtained over $\sim$6 h in late 2006. We observe enhanced emissions that appear close to latitudes of the UV northern aurora (delineated by \cite{Herbert:2009}). To confirm whether these emissions are auroral, the spectra were analysed for temperature, column density and total emissions to identify whether enhancements were thermally driven or created by an ion population increase.

\begin{figure}[h!]
\centering
\includegraphics[width=\linewidth]{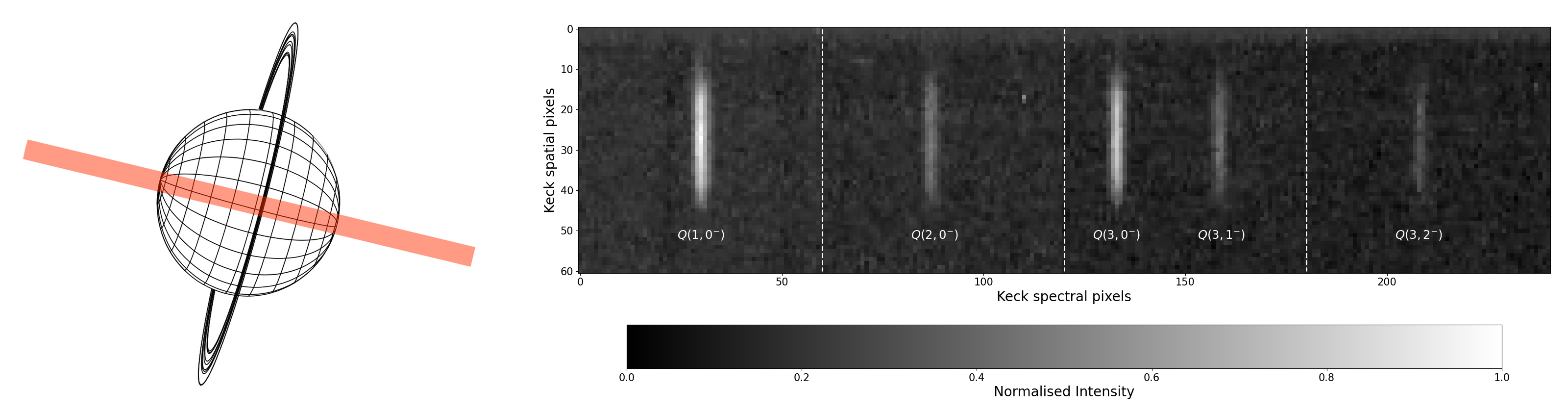}
\caption{\textbf{a}, The geometry of Uranus as was observed by KECK II SCAM (Slit-viewing camera) and NIRSPEC in September 2006. The orientation of the spectrometer slit on the disk of Uranus is shown in red, which aligns with the planet’s rotational poles. \textbf{b}, An averaged spectra obtained by KECK II NIRSPEC between 3.4 $\mu$m and 4.0 $\mu$m, including annotations for$\text{Q(1,0}^{\text{-}}\text{)}$, $\text{Q(2,0}^{\text{-}}\text{)}$, $\text{Q(3,0}^{\text{-}}\text{)}$, $\text{Q(3,1}^{\text{-}}\text{)}$ and $\text{Q(3,2}^{\text{-}}\text{)}$ emission lines, which can be found at 3.9530 $\mu$m, 3.9708 $\mu$m, 3.9860 $\mu$m, 3.9865 $\mu$m and 3.9946 $\mu$m, respectively. Panel a made in part with Uranus Viewer 3.1, Ring-Moon Systems Node (\url{https://pds-rings.seti.org/tools/viewer3$\_$ura.shtml}).}
\label{fig:1ab}
\end{figure}

Uranus observations were taken with the Keck II telescope on the 5th September 2006, from 07:26 to 13:24 UT, close to the planet’s equinox in 2007, using the NIRSPEC (Near-infrared Spectrograph) instrument \cite{McLean:1998} with a KL atmospheric window filter. A 0.288 × 24 arcsec slit was aligned with the plant’s rotational axis (shown in Fig. \ref{fig:1ab}a). Spectra were gathered between 3.5 $\mu$m and 4.1 $\mu$m where the fundamental Q-branch of $\text{H}_{\text{3}}^{\text{+}}$ emissions lies (shown in Fig. \ref{fig:1ab}b; raw image in Extended Data Figure 1). This triatomic hydrogen ion is a major constituent of Uranus’s ionosphere and planets whose upper atmosphere is dominated by molecular and ionic hydrogen \cite{Moore:2019}. A total of 218 spectra were taken over an ~6 h period with an integration time of ~30 s. These were co-added into 13 datasets to enhance the signal-to-noise ratio (total integration time per set was ~27 min). To increase the signal-to-noise ratio further, spatial pixels along the slit were grouped every 0.32 arcsecs (full details in the Methods). The exact longitude of Uranus has been completely lost; therefore, an arbitrary longitude has been selected for these results. Astronomical seeing on the night averaged at 0.44 arcsec, which is equivalent to a blur of 14\degree latitude and 12\degree longitude. During the observation, Uranus rotated by $\sim$180\degree and hence our final mapping spans an area up to $\sim$180\degree longitude. Unfortunately, a lapse in guiding between 10:52 UT and 11:31 UT resulted in the loss of 2 longitudinal data bins, leaving a gap in the middle of our scans. Finally, results presented here are not corrected for line of sight (LOS) (for example, see \cite{Johnson:2018}) and hence we expect infrared emissions to be enhanced near the planet’s limb. At Jupiter and Saturn, auroral emissions are LOS enhanced; however, Uranus’s solar extreme ultraviolet (EUV)-generated ionosphere is darker at the limbs \cite{Lamy:2018}, and, so, without a detailed understanding of the ionospheric brightening source, it is not possible to correct. However, as much of the enhanced emissions are away from the limbs, we expect minimal change in the location of emissions peaks after corrections.

\begin{figure}[h!]
\centering
\includegraphics[width=\linewidth]{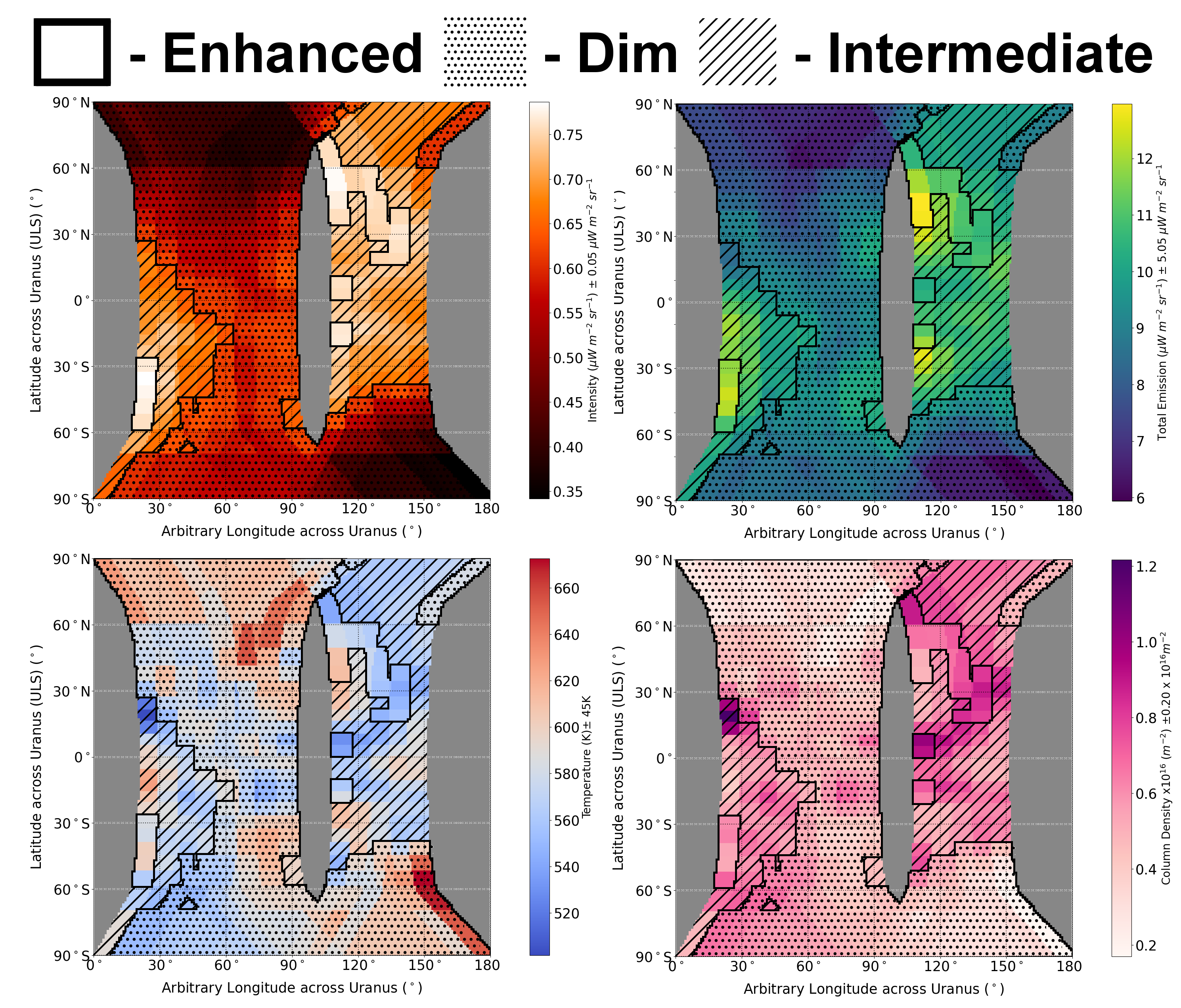}
\caption{\textbf{a}, Measured $\text{H}_{\text{3}}^{\text{+}}$ $\text{Q(1,0}^{\text{-}}\text{)}$ intensity mapped across the upper atmosphere of Uranus against Uranian latitude and arbitrary longitude. \textbf{b}, Total $\text{H}_{\text{3}}^{\text{+}}$ emission calculated from the temperature and column density (explained in detailed in Methods). \textbf{c}, Estimated temperatures of the $\text{H}_{\text{3}}^{\text{+}}$ emissions from all five Q-branch lines. \textbf{d}, Estimated column densities of $\text{H}_{\text{3}}^{\text{+}}$ emissions from all five Q-branch lines. The latitude is planetocentric whereas the longitude is arbitrary due to the loss of the ULS since Voyager II. The solid black lines mark out the boundaries of E1 (left) and E2 (right). Within the boundaries, the enhanced regions are unshaded, the dim regions are shaded with dots and the intermediate regions are shaded with diagonal lines.}
\label{fig:2abcd}
\end{figure}

To calculate the $\text{H}_{\text{3}}^{\text{+}}$ intensities, temperatures, column densities and total $\text{H}_{\text{3}}^{\text{+}}$ emission for the upper atmosphere of Uranus, this study focuses on five quasi-thermalized ro-vibrational emission lines of $\text{H}_{\text{3}}^{\text{+}}$, $\text{Q(1,0}^{\text{-}}\text{)}$, $\text{Q(2,0}^{\text{-}}\text{)}$, $\text{Q(3,0}^{\text{-}}\text{)}$, $\text{Q(3,1}^{\text{-}}\text{)}$ and $\text{Q(3,2}^{\text{-}}\text{)}$; these physical parameters were calculated from a full spectrum best fit, as described in Methods. The final fitted spectra provide intensity values that are then mapped across Uranus as is shown with the $\text{Q(1,0}^{\text{-}}\text{)}$ emission line (with the highest signal-to-noise ratio; Fig. \ref{fig:2abcd}a), $\text{H}_{\text{3}}^{\text{+}}$ total emission (Fig. \ref{fig:2abcd}b), temperatures (Fig. \ref{fig:2abcd}c) and $\text{H}_{\text{3}}^{\text{+}}$ column density (Fig. \ref{fig:2abcd}d). The respective error maps are shown in Extended Data Figure 2.

In Fig. \ref{fig:2abcd}a, b, the $\text{H}_{\text{3}}^{\text{+}}$ emission intensity varies with local time. To confirm the source of these enhancements, we define three regions of interest that are algorithmically distinct: the ‘enhanced’ region where the emissions are brighter than the mean plus one standard deviation (shown in solid black lines but not shaded); the ‘dim’ region where emissions are below the mean emission (shaded with dots); and the ‘intermediate’ region where emissions are brighter than the mean, but within a standard deviation of that mean (shaded by diagonal lines). The means and standard deviations for Fig. \ref{fig:2abcd} presented in Table \ref{tab:1} are the result of subtracting each pixel by its uncertainty (seen in Extended Data Figure 2). The resulting datasets are hence minimized, meaning pixels in the enhanced region are statistically significant.

\begin{table}[ht]
\centering
\begin{tabular}{|l|l|l|l|l|}
\hline
 & Mean $\text{Q(1,0}^{\text{-}}\text{)}$ intensity & Mean temperature & Mean column density & Total $\text{H}_{\text{3}}^{\text{+}}$ emission \\
  & ($\mu$W$\text{m}^{\text{-2}}$$\text{sr}^{\text{-1}}$) & (K) & ($\times$ $\text{10}^{\text{15}}$ $\text{m}^{\text{-2}}$) & ($\mu$W$\text{m}^{\text{-2}}$$\text{sr}^{\text{-1}}$) \\
\hline
Enhanced E1 & 0.723 $\pm$ 0.010 & 585 $\pm$ 14 & 4.017 $\pm$ 0.457 & 6.155 $\pm$ 0.887 \\
\hline
Enhanced E2 & 0.716 $\pm$ 0.009 & 564 $\pm$ 22 & 5.113 $\pm$ 0.826 & 6.354 $\pm$ 0.616 \\
\hline
Intermediate & 0.650 $\pm$ 0.032 & 572 $\pm$ 19 & 4.134 $\pm$ 0.761 & 5.354 $\pm$ 0.664 \\
\hline
Dim & 0.472 $\pm$ 0.086 & 593 $\pm$ 24 & 2.432 $\pm$ 0.901 & 3.212 $\pm$ 1.235 \\
\hline
\end{tabular}
\caption{\label{tab:1} \textbf{Means and standard deviations of the $\text{Q(1,0}^{\text{-}}\text{)}$ intensities, $\text{H}_{\text{3}}^{\text{+}}$ ion temperatures, ion column densities and total emission for the enhanced and dim regions}}
\end{table}

In Fig. \ref{fig:2abcd}a, the enhanced regions show intriguing structures, the first, which is smaller, between 26\degree S and 59\degree S and from 18\degree to 28\degree longitude (E1). The second area extends between 15\degree N and 75\degree N from 100\degree and 143\degree longitude with two smaller emission spots between 10\degree N and 0\degree and between 10\degree S and 20\degree S over a 108\degree to 117\degree  longitude range, which we refer to as E2. Table 1 summarizes the mean values of $\text{Q(1,0}^{\text{-}}\text{)}$ intensities, along with mean values for temperature, column density and total $\text{Q(1,0}^{\text{-}}\text{)}$ emission. Comparing the dim region’s mean $\text{Q(1,0}^{\text{-}}\text{)}$ intensity (0.472 $\pm$ 0.086 $\mu$W$\text{m}^{\text{-2}}$$\text{sr}^{\text{-1}}$) with that of the enhanced region (0.723 $\pm$ 0.010 $\mu$W$\text{m}^{\text{-2}}$$\text{sr}^{\text{-1}}$ and 0.716 $\pm$ 0.009 $\mu$W$\text{m}^{\text{-2}}$$\text{sr}^{\text{-1}}$, we find a 27 \% to 90 \% enhancement.

Figure \ref{fig:2abcd}b shows the total $\text{H}_{\text{3}}^{\text{+}}$ emission, which is the combined intensity from all $\text{H}_{\text{3}}^{\text{+}}$ emission lines in this investigation. We find the two enhanced regions average at 6.155 $\pm$ 0.681 $\mu$W$\text{m}^{\text{-2}}$$\text{sr}^{\text{-1}}$ and 6.354 $\pm$ 0.616 $\mu$W$\text{m}^{\text{-2}}$$\text{sr}^{\text{-1}}$ for E1 and E2, respectively, while the dim region emits at a lower average of 3.212 $\pm$ 1.235 $\mu$W$\text{m}^{\text{-2}}$$\text{sr}^{\text{-1}}$, hence an 18 \%, up to 353 \% increase at both E1 and E2. This large range in emission enhancement is most likely from the high uncertainty in column density, which affects the error propagation when calculating the total emission. We, however, conclude that our division of emissions of Uranus—whether the single $\text{Q(1,0}^{\text{-}}\text{)}$ line or total $\text{H}_{\text{3}}^{\text{+}}$ emission, into distinct enhancement related regions, is both robust and significant.

Comparing physical parameters between enhanced regions provides an understanding of how they are enhanced. The average temperature for the dataset is 585 $\pm$ 25 K, which aligns with previous temperature observations (\cite{Melin:2011} for 2006 at 608 $\pm$ 12 K). The enhanced regions have a mean temperature of 585 $\pm$ 14 K and 564 $\pm$ 22 K for E1 and E2, respectively, with the dim region temperature at 593 $\pm$ 24 K, shown in Fig. 2c. While the enhanced regions appear cooler, there is overlap in temperature errors, so while we cannot conclude the emission is anticorrelated with temperature, thermal processes cannot explain the intensity enhancements.

Except at the planet’s limbs, EUV ionization produces a uniform column ionization rate across the whole disk; however, enhancements of column densities could be produced by enhanced particle precipitation, suggestive of auroral activity. Shown in Fig. \ref{fig:2abcd}d, we observe an average column-density difference of 2.133 $\times$ $\text{10}^{\text{15}}$ $\text{m}^{\text{-2}}$ at the enhanced regions (4.017 $\pm$  0.457 $\times$ $\text{10}^{\text{15}}$ $\text{m}^{\text{-2}}$ and 5.113 $\pm$ 0.826 $\times$ $\text{10}^{\text{15}}$ $\text{m}^{\text{-2}}$ at E1 and E2, respectively) compared with the dim region (2.432 $\pm$ 0.901 $\times$ $\text{10}^{\text{15}}$ $\text{m}^{\text{-2}}$). These densities, on average, are higher (about two to five times higher) than reported in \cite{Melin:2011}. Here a more through and complete data-reduction process was conducted over the whole night of observations rather than half the night, with densities presented in Table 1 aligning within the range of densities observed in previous investigations \cite{Melin:2011},\cite{Melin:2019}. In contrast to the temperatures, the enhanced region’s column density is on average 88\% enhanced. Put simply, more emitters, rather than hotter emitters, is what is resulting in more emission.

There are several scenarios that could lead to a column-density enhancement at locations of increased $\text{H}_{\text{3}}^{\text{+}}$ emissions. One possibility is if the ion is produced evenly across the planetary disk, there is some mechanism by which it is transported from the dim region into the enhanced regions.

We do not consider meridional transport from the rotational poles to be substantial for two reasons: first, Uranus is a large, rapidly rotating planet where it is difficult to overcome the Coriolis forces; second, if there are equatorwards winds, we would expect to see a $\text{H}_{\text{3}}^{\text{+}}$ bulge evenly distributed at lower latitudes. There is nothing in our data to suggest polewards meridional transport.

Zonal winds on Uranus are generally between 0 and 250 $\text{ms}^{\text{-1}}$. A previous study \cite{Lindal:1987} found electron densities between $\approx$ $\text{10}^{\text{9}}$ $\text{m}^{\text{-3}}$ (Voyager egress) and $\approx$ $\text{10}^{\text{10}}$/~$\text{10}^{\text{11}}$ $\text{m}^{\text{-3}}$ (Voyager ingress). Taking these figures together with the dissociative recombination coefficient of $\approx$ $\text{10}^{\text{-13}}$ $\text{m}^{\text{3}}$$\text{s}^{\text{-1}}$ (\cite{Larsson:2008}) suggests a maximum half lifetime $\tau$($\text{H}_{\text{3}}^{\text{+}}$) of less than $\text{10}^{\text{4}}$s, and possibly as low as 100 s. Hence an individual $\text{H}_{\text{3}}^{\text{+}}$ ion could be transported $\approx$ 2,000 km. This is less than the ~30,000 km at the equator to get from the centre of E2; although the distance from there to the centre of E1 is approximately half that value, it is still too far. Hence, we assume that the $\text{H}_{\text{3}}^{\text{+}}$ ions, their emissions and physical parameters are representative of locally produced features.

\begin{figure}[h!]
\centering
\includegraphics[width=\linewidth]{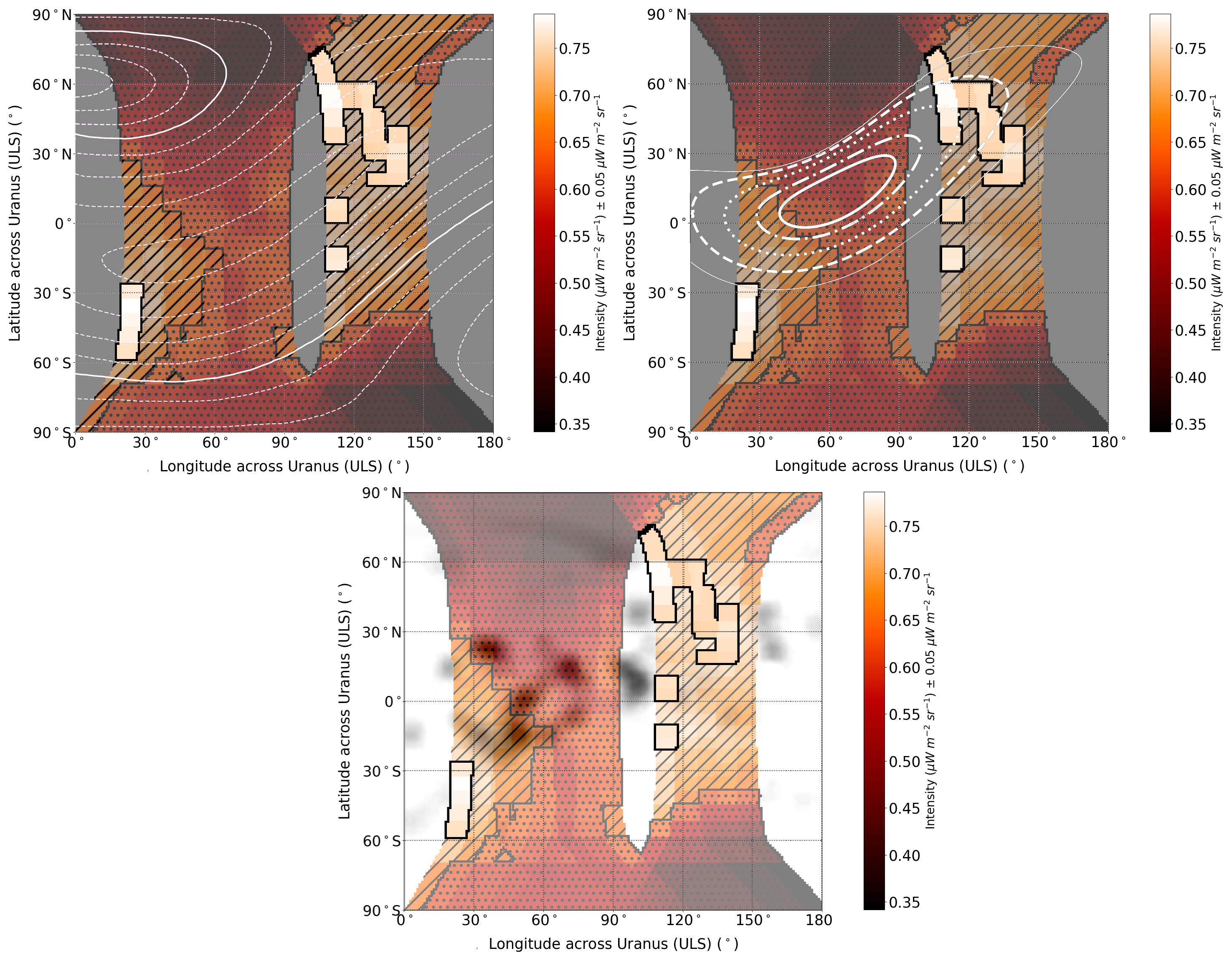}
\caption{\textbf{a}, $\text{Q(1,0}^{\text{-}}\text{)}$ mapped intensities with the $\text{Q}_{\text{3}}$ model with contours representing the contours of 20° dip angles to the thick continuous white line, which is the magnetic-field equator. Here the ULS has been sourced and placed into our observations from the $\text{Q}_{\text{3}}$ model. The grey background colour represents the areas that were unobserved in these observations. \textbf{b}, $\text{Q(1,0}^{\text{-}}\text{)}$ intensities (where only the enhanced region has been highlighted) mapped alongside the L shells of 2 (solid line), 3 (dashed), 5 (dotted), 10 (dot and dashed) and 20 (solid) of the $\text{Q}_{\text{3}}^{\text{mp}}$ model. \textbf{c}, $\text{Q(1,0}^{\text{-}}\text{)}$ intensities (where only the enhanced region has been highlighted) mapped alongside the H2 band emissions intensity map from \cite{Herbert:2009} To avoid obscuring the UV emissions, the dark grey background has been removed in this panel.}
\label{fig:3abc}
\end{figure}

Another potential driver for the dim region’s low column densities could be ‘ring rain’ as seen at Saturn \cite{ODonoghue:2013}. Here, $\text{H}_{\text{3}}^{\text{+}}$ destruction is modulated by water molecules in the planet’s rings travelling along the field lines into the planet’s lower latitudes. Fig. \ref{fig:3abc}a combines Fig. \ref{fig:2abcd}a and the $\text{Q}_{\text{3}}$ model from \cite{Connerney:1987}, which maps Uranus’s magnetic field with dip angle contours, using contour steps of 20 \degree dip angle (the angle made with the planet’s horizontal plane by its magnetic-field lines). We expect the ring rain to affect only a narrow band of dip angles (mapping to the planet’s rings, 1.6–2 $\text{R}_{\text{U}}$), where in Fig. \ref{fig:3abc}a we observe the dim region over a large range of dip angles. Hence, quenching ring rain cannot explain the emissions we observe.

Two more magnetic-field models $\text{Q}_{\text{3}}^{\text{mp}}$ and $\text{AH}_{\text{5}}$; \cite{Herbert:2009}) have since been used at Uranus, replacing previous models with a more globally representative magnetopause image field and including UV auroral emissions from Voyager II, respectively. These models provide a strong fit to the southern aurora, but the northern aurora is poorly constrained as Voyager crossed magnetic-field lines that mapped close to the southern magnetic pole twice, once at a distance of 4.19 $\text{R}_{\text{U}}$, but only once at the north, at >20 $\text{R}_{\text{U}}$. In addition, the auroral morphology may have changed with solar-wind pressure or by changes in the preferred auroral acceleration region above the planet. Given this complexity, we focus solely on the $\text{Q}_{\text{3}}$ model.

As none of the previous processes can explain the NIR enhancement morphology, the most plausible explanation is that the density enhancements are driven by auroral production. In previous $\text{H}_{\text{3}}^{\text{+}}$ investigations at Jupiter and Saturn \cite{Drossart:1989},\cite{Geballe:1993},\cite{Stallard:2008},\cite{Stallard:2019},\cite{Johnson:2018},\cite{Chowdhury:2019}, we have observed the strongest infrared emission enhancements and column densities at the auroral regions, where particle precipitation results in significant ionization in the upper atmosphere. We find that the enhanced regions strongly suggest auroral production and so consider that we have partially mapped the northern infrared aurora.

To determine how the 2006 infrared emissions aligned with previous models, we have chosen to not add a longitude shift into our work due to the lack of known longitude (ULS) in 2006. Comparing Fig. \ref{fig:2abcd}a and the $\text{Q}_{\text{3}}$ model (Fig. \ref{fig:3abc}a), we observe intensities between 60 \degree and 80 \degree dip angle at the same angle as where the auroral oval sits, although the approximate location of the auroral oval in the $\text{Q}_{\text{3}}$ model sits within the dim region (similar in location to Fig. \ref{fig:3abc}b). Other enhanced regions with poor alignment (where the dip angle drops to 20 \degree) may be due to more complex morphology within the surface magnetic-field structure, or the effects of seeing (at least $\approx$ $\pm$12 \degree) along with the low spatial resolution ($\approx$ 0.32 arcsec). We also note relatively weak emissions between 40 \degree and 100 \degree longitude. While most pixels in this region are ‘dim’, we highlight that these emissions remain brighter than the limbs. It may be that while the enhancement is not significant as the enhanced region, only the edges of the map represent the EUV ionized background $\text{H}_{\text{3}}^{\text{+}}$ density. This weaker central region could be driven by weaker auroral precipitation and hence further investigations are required.

In Fig. \ref{fig:3abc}b, we compare the infrared emissions against the L-shell magnetic-field lines of the $\text{Q}_{\text{3}}$ model. Here the emissions observed in both E1 and E2 extend out past the optimal L5 shell, which is where the brightest UV emissions are observed by Voyager II and HST. Focusing on E1, between 30\degree S and 60\degree S and before 30\degree ULS longitude, we find no enhanced emissions align with the $\text{Q}_{\text{3}}$ L shells. These emissions are, however, located close to dayside O-source radio emissions \cite{Desch:1991}, and are close to n-smooth radio emissions observed in \cite{Kaiser:1989}, where the authors suggested that these emissions arose from unusual particle distribution from particle absorption by the $\epsilon$ ring, which may act as a driver for these infrared extended emissions. We do, however, find a portion of E2 emissions fit within L shells of 3 and 5, where weak UV emissions in Fig. \ref{fig:3abc}c are located ($\approx$ <100 R).

Figure \ref{fig:3abc}c compares Fig. \ref{fig:2abcd}a with UV auroral emissions from Voyager II in 1986. At Jupiter, UV and infrared aurora appear at similar latitudes \cite{Clarke:2004}, \cite{Melin:2006}; however, at the auroral oval, \cite{Radioti:2013} found UV and infrared auroral features’ brightness can vary independently of each other and hence are not co-located. We should then not expect the brightest NIR emissions to be co-located with the brightest UV emissions at Uranus. Further differences between infrared and UV emissions can also be explained by the $\approx$ 15 min lifetime of $\text{H}_{\text{3}}^{\text{+}}$ (at Jupiter) smoothing out short-term (1–2 min) variability in UV emissions \cite{Stallard:2016}. A similar effect may also occur at Uranus.

The enhanced $\text{H}_{\text{3}}^{\text{+}}$ emissions are broadly spaced in latitude compared with the brightest UV emission, where the strongest UV emissions occur north of E1 and only weaker UV emissions appear close to or at E2. This spreading of infrared emissions suggests that $\text{H}_{\text{3}}^{\text{+}}$ emissions occur more equatorwards (magnetic-field equator) than the UV emissions and appear anticorrelated in terms of longitude. Differences in emission region may result from changes in the auroral drivers, changes in the solar wind (as observed at Jupiter and Saturn \cite{Clarke:2009}); or short-term variability associated with the local time. These might be via changing precipitation flux or precipitation energy. Equally, contrasts in apparent magnetic mapping of the two aurorae could originate from poor alignment of our arbitrary longitude. It is difficult to draw too many conclusions without re-discovering the rotational phase of Uranus.

This likely detection of $\text{H}_{\text{3}}^{\text{+}}$ aurora at Uranus has broader implications for Neptune, given the planet’s similarities (similar unaligned and offset magnetic fields). Currently, we have not detected $\text{H}_{\text{3}}^{\text{+}}$ at Neptune \cite{Melin:2018}, the cause suggested to be a cooler than expected upper atmosphere \cite{Moore:2019}. The presence of infrared aurora at Uranus suggests the potential for detecting aurora at Neptune, where past observations may have been taken during weak emission periods.

Confirming infrared aurora at Uranus directly assists in exoplanetary advancements as warm-ice-giant worlds make up a large fraction of the current population \cite{Wolfgang:2016}. Auroral detections from exoplanets could provide upper atmospheric diagnostics. A detailed understanding of Uranus could advance exoplanet knowledge, helping the wider scientific community to understand their ionospheric compositions. Uranus also presents an ideal laboratory for observing conditions during magnetic-field reversal, as the magnetospheric axis consistently changes direction with respect to the solar wind, over a single Uranian day \cite{Cowley:2013}. Current research requires analysis of volcanic rock at Earth or modelling to identify the effects of a reversal \cite{Glatzmaier:1995}. By undertaking consistent observations, we can identify changes in atmospheric processes, which extrapolated to Earth can enhance our modelling of misaligned magnetic fields.

\section*{Methods}
\subsection*{Introduction}

This study uses 216 spectral images of $\text{H}_{\text{3}}^{\text{+}}$ emission taken on the 5th September 2006, the year before Uranus would reach equinox. These images were taken using the NIRSPEC instrument at the W.M. Keck II Observatory using a KL atmospheric window filter that covers the 3.3–5.0 $\mu$m range, reduced through the spectrometer grating to a 3.3 5– 4.0 $\mu$m range to focus on the fundamental Q-branch emission of $\text{H}_{\text{3}}^{\text{+}}$ (which is known from previous investigations \cite{Melin:2011},\cite{Melin:2019} to emit with a suitable signal-to-noise ratio). This wavelength range was used across all images including the reference star (HR 215 143) and flat-field images.

Due to the slit length of 24'' across Uranus (which at the time subtended 3.7'' in the sky), we positioned the disk of the planet at two locations along the slit when taking images. The first frame (known as the A frame) centred the disk of Uranus in the top half of the slit, while the second frame (known as the B frame) centred the disk of Uranus in the bottom half of the slit. By carrying out this pattern, for the same sky position we would have alternating observations of Uranus’s and Earth’s atmospheric emission and the second with just Earth’s atmospheric emission. The latter data could then be subtracted from the former to mitigate the effect of sky emissions.

Once all images were obtained, the data were then reduced using the Interactive Data Language (IDL)-based RedSpec data-reduction pipeline, which shifts data into straightened two-dimensional arrays by using the reference star images, as shown in the reference material. To mitigate the effect of thermal currents or dead pixels on the detector, all reduced images were then calibrated using the flat and dark frames.

\subsection*{$\text{H}_{\text{3}}^{\text{+}}$ intensity calculations and mapping}

Taking the images from the reference star frames (which at the time was closest to Uranus within the night sky and shown in Supplementary Fig. 1), the flux can be estimated by taking a Gaussian fit across the blackbody emission lines (similar to the Gaussian fit shown in Supplementary Fig. 2a,b). Once found, the measured flux observed can be compared against the known intensity of HR 215 143, which is estimated using the work of \cite{Blackwell:1983} with an $\alpha$ Lyrae (an A0) star. In this work, HR 215 143 is a B7.5 V star, although the radius, mass and luminosity are of the same or similar magnitude to that of A0 (A0V) stars \cite{Pecaut:2013}; hence, we use the calculations of \cite{Blackwell:1983} as a close representation for the expected flux of HR 215 143. The expected temperature of HR 215 143 is, however, at approximately 11,000 K instead of 10,000 K (as expected of A0 stars) as calculated from the work of \cite{Silaj:2014}; this difference in temperature is included in the calculations described above.

This ratio can then be used to convert detector counts from images into a known intensity value, which was carried out across all images. These images were then grouped into 54 sets of data and aligned (in case Uranus had appeared to shift across the slit) by using a Python script that detected the disk of the planet by measuring the central position within the longest sequence of pixels with emission values 1 s.d. greater than the background emission and shifting data by the required number of pixels so that individual spectra could be co-added.

When all the data had been correctly lined up, to enhance the signal-to-noise ratio, the data were binned into 13 temporal sets of data across ~6 h of observations. This was completed by the sequence in which A frames and B frames were taken, in an ABBA pattern where two added A frames are subtracted by two B frames and averaged. By averaging four datasets (over the previous 54 sets) at a time, it was possible to obtain a spectrum with sufficient signal-to-noise ratio to complete Gaussian fitting, as seen in Supplementary Data Fig. 2a,b. Originally Uranus was found to cover just under 23 pixels (22.9 pixels) across the detector, the limbs of which were found by code that searched each of the 13 temporal datasets for enhanced emission lines that extended for 22–23 pixels; from this the approximate middle of the disk could be deduced.

A further enhancement of the signal-to-noise ratio was required to minimize the error in intensity and so a 2 pixel weighted rolling average was chosen to enhance the signal strength. This would mean starting from the northern limb, the first and second pixels were averaged and assigned to the first mapped pixel, then the second and third pixels were averaged and assigned the second mapped pixel and so on. This resulted in a total of 22 pixels across Uranus, allowing identification of auroral or enhanced regions.

To find the observed intensity from all emission lines, the h3ppy Python package (a Python version of the C++ fitting procedure as detailed in \cite{Melin:2006}) was used. This produces multiple Gaussian fits across a spectrum with a known range of wavelength, with intensities for each line varying with modelled temperature and column density. The calculations behind this are explained in detail in \cite{Johnson:2017}. It should be noted due to the changing LOS of the observer across Uranus’s disk, the intensity will be enhanced at the limb of the planet due to the observer effectively viewing through more of the atmosphere. The background solar EUV when modelled at Uranus diminishes at the edges and with the majority of the data localized away from the limb, we have not adjusted for this effect. The errors in the intensity of $\text{Q(1,0}^{\text{-}}\text{)}$ were calculated by the errors in fitting a Gaussian curve with the observed emission line (errors predominately arising from the height and width of the fit).

\subsection*{$\text{H}_{\text{3}}^{\text{+}}$ ro-vibrational temperature calculations}

By applying a full spectra best fit across emission lines $\text{Q(1,0}^{\text{-}}\text{)}$, $\text{Q(2,0}^{\text{-}}\text{)}$, $\text{Q(3,0}^{\text{-}}\text{)}$, $\text{Q(3,1}^{\text{-}}\text{)}$ and $\text{Q(3,2}^{\text{-}}\text{)}$, the ro-vibrational temperatures can be calculated using ab initio Einstein A coefficients—physical parameters of $\text{H}_{\text{3}}^{\text{+}}$ spectra emission lines as detailed by \cite{Neale:1996} and the upper energy levels as described by \cite{Miller:1989}. This process is carried out with the h3ppy package (see Supplementary Fig. 3 for a visual representation of this fitting with an averaged spectrum from the 5 September) assuming a quasi-local thermodynamic equilibrium $\text{H}_{\text{3}}^{\text{+}}$ spectra fit20.

\subsection*{$\text{H}_{\text{3}}^{\text{+}}$ column-density calculations}

This data product was calculated by dividing the measured intensity from the observed emission lines by the theoretical emission per molecule for all the Q-branch emission lines mentioned in the temperature calculations, as described by \cite{Trafton:1999}. It should be noted due to the LOS of the observer across Uranus’s disk, the column density will be enhanced at the limb of the planet due to the observer effectively viewing through the atmosphere twice. At present, these results have not been adjusted for this effect, as most of the data are localized away from the limb of the planet, but we estimate that, within the auroral regions discussed, this enhancement would be 14 \%.

\subsection*{$\text{H}_{\text{3}}^{\text{+}}$ total emission calculations}

Using the calculations of \cite{Neale:1996}, the total emission can be calculated by the product of the number of ions by the temperature-dependent total emission per molecule (Emol) while assuming local thermal equilibrium. This requires both the column density and temperature over two or more emission lines), where temperature is used to calculate Emol. It should be highlighted that due to temperatures staying between 500 K $\leq$ T $\geq$ 900 K, suitable coefficient values were selected to calculate Emol.

\section*{Data availability}

The NIRSPEC raw data used in this study (and subsequent raw data used in Figs. \ref{fig:1ab}, \ref{fig:2abcd} and \ref{fig:3abc}) are publicly available on the Keck Observatory Archive (KOA) at \url{https://koa.ipac.caltech.edu/cgi-bin/KOA/nph-KOAlogin} and included with the source data. Reduced and calibrated images used in this current study can be obtained through the RedSpec code (discussed below) with the final data used in the figures of this paper available with the source data and \url{https://github.com/physicist-et/Uranus_AuroraKeck_0905}. Source data are provided with this paper.

\section*{Code availability}

RedSpec is a data-reduction package in IDL, designed to reduce and process spectral images from NIRSPEC and is available at \url{https://www2.keck.hawaii.edu/inst/nirspec/redspec}. h3ppy is a $\text{H}_{\text{3}}^{\text{+}}$ emission modelling and fitting package in Python and is available at \url{https://github.com/henrikmelin/h3ppy}. All remaining code used to extract the reduced data, align it for use with h3ppy and Gaussian fit function and mapping variables are available at \url{https://github.com/physicist-et/Uranus_AuroraKeck_0905}.

\printbibliography

\section*{Acknowledgements}

The UK STFC studentships ST/T506242/1 and ST/N504117/1 supported the work of E.M.T. and M.N.C., respectively. The STFC James Webb Fellowship (ST/W001527/1) at the University of Leicester, UK supported H.M., and the UK STFC Consolidated Grant ST/N000749/1 supported T.S.S. The data presented herein were obtained at the W. M. Keck Observatory, which is operated as a scientific partnership among the California Institute of Technology, the University of California and the National Aeronautics and Space Administration. The observatory was made possible by the generous financial support of the W. M. Keck Foundation. The authors are indebted to Mark Showalter for their assistance with Fig. \ref{fig:1ab} of this article. The authors recognize and acknowledge the very significant cultural role and reverence that the summit of Maunakea has always had within the Indigenous Hawaiian community. We are most fortunate to have the opportunity to conduct observations from this mountain.

\section*{Author information}
\subsection*{Authors and Affiliations}

\textbf{School of Physics and Astronomy, University of Leicester, Leicester, UK}
\newline Emma M. Thomas, Henrik Melin, Mohammad N. Chowdhury \& Ruoyan Wang

\textbf{Department of Mathematics, Physics and Electrical Engineering, Northumbria University, Newcastle upon Tyne, UK}
\newline Tom S. Stallard \& Katie Knowles

\textbf{Department of Physics and Astronomy, University College London, London, UK}
\newline Steve Miller

\subsection*{Contributions}
E.M.T. performed data reduction and data analysis, and contributed to the writing and editing of the paper. H.M. was responsible for the data analysis with the use of h3ppy. T.S.S. contributed via data reduction and data analysis. M.N.C. was responsible for IDL to Python code conversion for two-dimensional mapping over a three-dimensional surface. K.K. and R.W. contributed to the discussion and editing of the paper. S.M. is the principal investigator for the observations of Uranus taken in 2006 and contributed to the paper via discussion and editing of the paper.

\subsection*{Corresponding author}
Correspondence to Emma M. Thomas.

\section*{Ethics declarations}
\subsection*{Competing interests}
The authors declare no competing interests.

\section*{Peer review}
\subsection*{Peer review information}
\textit{Nature Astronomy} thanks James Sinclair and the other, anonymous reviewer(s) for their contribution to the peer review of this work.

\section*{Additional information}
\textbf{Publisher's note} Springer Nature remains neutral with regard to jurisdictional claims in published maps and institutional affliations.

\section*{Extended Data}
\begin{figure}[h!]
\centering
\includegraphics[width=\linewidth]{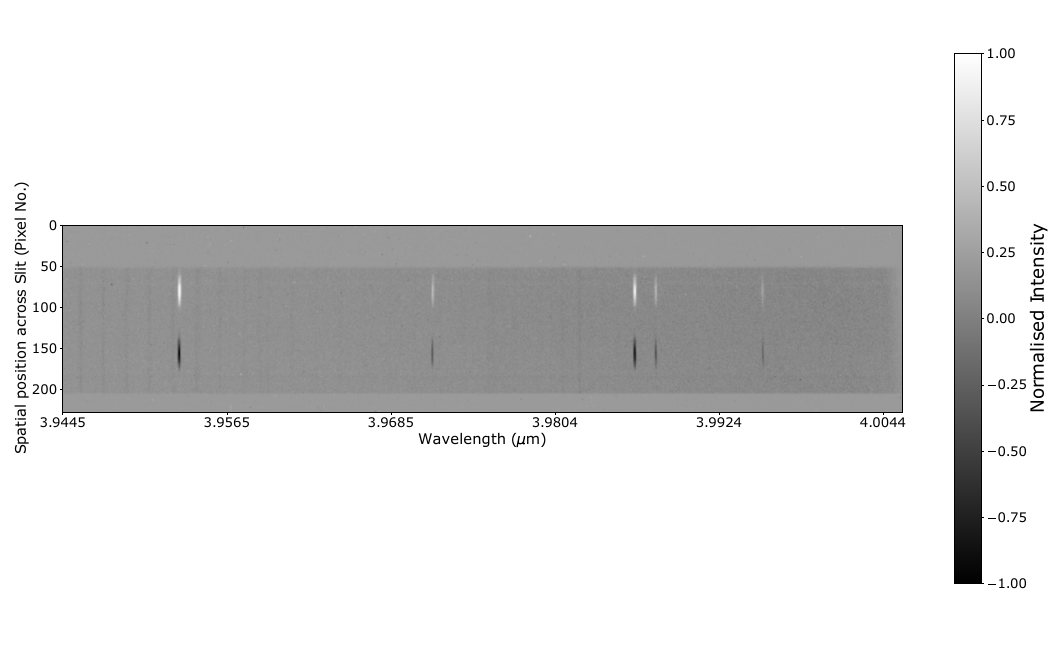}
\caption{\textbf{Extended Data Fig. 1 Averaged ABBA spectrum of Uranus observed between 3.9445 $\mu$m and 4.0044 $\mu$m by NIRSPEC on the 5th September 2006.} The $\text{Q(1,0}^{\text{-}}\text{)}$ emission line can be observed at 3.9530 $\mu$m with the $\text{Q(2,0}^{\text{-}}\text{)}$, $\text{Q(3,0}^{\text{-}}\text{)}$, $\text{Q(3,1}^{\text{-}}\text{)}$ and $\text{Q(3,2}^{\text{-}}\text{)}$ emission lines at 3.9708, 3.9860, 3.9865 and 3.9946 $\mu$m respectively.}
\label{fig:EX1}
\end{figure}

\begin{figure}[h!]
\centering
\includegraphics[width=\linewidth]{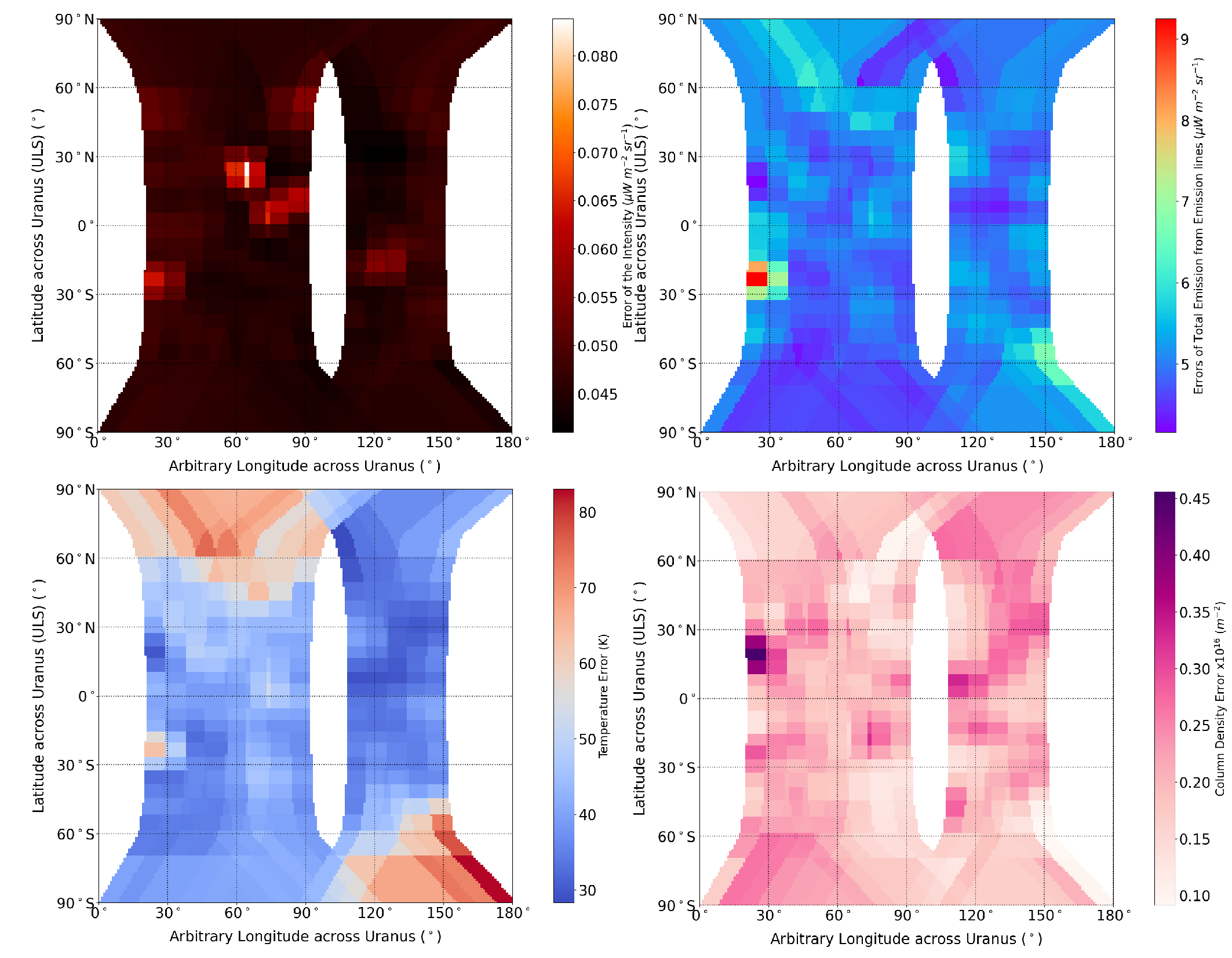}
\caption{\textbf{Extended Data Fig. 2 a) Intensity errors calculated from the $\text{Q(1,0}^{\text{-}}\text{)}$ emission line, b) Total emission errors c) Temperature errors and d) Column Density errors across 180° longitude.}
Errors for the Intensity were calculated from the error in fitting the Gaussian with the emission line via the error in height and width propagated. Errors in Total Emission, Temperature and Column Density were calculated through h3ppy’s fit of the emission spectra and following calculations.}
\label{fig:EX2}
\end{figure}

\section*{Supplementary information}
\subsection*{Supplementary Information}
Supplementary Figs. 1 - 3. \url{https://static-content.springer.com/esm/art%3A10.1038%2Fs41550-023-02096-5/MediaObjects/41550_2023_2096_MOESM1_ESM.pdf}

\subsection*{Supplementary Data 1}
Contains error mapping for the H3+ spectrum from Uranus of intensity, total emission, temperature and column densities. \url{https://static-content.springer.com/esm/art%3A10.1038%2Fs41550-023-02096-5/MediaObjects/41550_2023_2096_MOESM2_ESM.xlsx}

\section*{Source data}
\subsection*{Source Data Fig.1}
Reduced and averaged Uranus spectra over the 3.94 $\mu$m to 4.04 $\mu$m wavelength. \url{https://static-content.springer.com/esm/art%3A10.1038%2Fs41550-023-02096-5/MediaObjects/41550_2023_2096_MOESM3_ESM.xlsx}

\subsection*{Source Data Fig. 2}
Reduced and mapped intensities, total emission, temperature and column densities for observations from Keck II NIRSPEC on the 5th September 2006. \url{https://static-content.springer.com/esm/art%3A10.1038%2Fs41550-023-02096-5/MediaObjects/41550_2023_2096_MOESM4_ESM.xlsx}

\section*{Rights and permissions}
\textbf{Open Access} This article is licensed under a Creative Commons Attribution 4.0 International License, which permits use, sharing, adaptation, distribution and reproduction in any medium or format, as long as you give appropriate credit to the original author(s) and the source, provide a link to the Creative Commons license, and indicate if changes were made. The images or other third party material in this article are included in the article’s Creative Commons license, unless indicated otherwise in a credit line to the material. If material is not included in the article’s Creative Commons license and your intended use is not permitted by statutory regulation or exceeds the permitted use, you will need to obtain permission directly from the copyright holder. To view a copy of this license, visit \url{http://creativecommons.org/licenses/by/4.0/}.



\end{document}